\documentclass[aps,twocolumn,showpacs,groupedaddress]{revtex4}
\usepackage{graphicx}

\begin{document}

%\draft
\title{
Reply to 'comment on Berry phase in a composite system'}
\author{X. X. Yi$^1$, L. C. Wang $^1$, and T. Y. Zheng $^2$}
\affiliation{$^1$Department of Physics, Dalian University of
Technology, Dalian 116024, China\\
$^2$School of Physics, Northeast Normal University, Changchun
130024,China\\}

\date{\today}

 \maketitle

In their comment\cite{han} on our Letter\cite{yi04}, Han and Guo
point out that we did not obtain quantitatively correct results for
the Berry phase in the composite system, and the proposed subsystem
Berry phase is not well-defined. Then they present a calculation for
the Berry phase in the weak coupling limit $g\rightarrow 0$ and
discuss the definition of the subsystem geometric phase. While the
eigenstate $|\Psi_{1,3}(\phi, \theta, g)\rangle$ in the limit $ g
\rightarrow 0$ may be correct in their comment, the Berry phase is
not the result as  Han {\it etal.} presented in their comment for a
realistic composite system. As we will show, the contribution of the
second subsystem( the subsystem does not feel the magnetic fields)
to the Berry phase in the composite system tends to zero in the
limit $g \rightarrow 0$ in the system, and the geometric phase for
the subsystem is well defined.

We now turn to the details. The composite system is degenerate at
$g=0$. Thus we need to study the non-Abelian geometric phase at this
point. Take the eigenvalue $E_1$ as an example, the two degenerate
orthogonal eigenstates at $g=0$ are\cite{note1},
$|\Psi_{a,b}\rangle=(\cos\frac{\theta}{2}e^{-i\phi}|\uparrow\rangle+
\sin\frac{\theta}{2}|\downarrow\rangle)_1\otimes\frac{1}{\sqrt{2}}(|\downarrow\rangle\pm
e^{i\phi}|\uparrow\rangle)_2.$ Define $A_{xy}=i\langle
\Psi_x|\frac{\partial}{\partial \phi}|\Psi_y\rangle,$ $x,y=a,b$, we
obtain
\begin{eqnarray}
A= \left( \matrix{ \cos^2\frac{\theta}{2}-\frac 1 2 & \frac 1 2
  \cr
  \frac 1 2& \cos^2\frac{\theta}{2}-\frac 1 2
   \cr } \right).\nonumber\\
   \label{A}
\end{eqnarray}
As the non-Abelian geometric phase is not a gauge invariant
quantity, one can not measure all the matrix elements of $U_g={\cal
P} exp(i\oint  A d\phi),$ except the eigenvalues of $U_g$ and its
trace, which in this case contain no contribution from the second
subsystem. It is important to note that $
|\Psi_a^{'}\rangle=(\cos\frac{\theta}{2}e^{-i\phi}|\uparrow\rangle+
\sin\frac{\theta}{2}|\downarrow\rangle)_1\otimes |\uparrow\rangle_2
$ and
$|\Psi_b^{'}\rangle=(\cos\frac{\theta}{2}e^{-i\phi}|\uparrow\rangle+
\sin\frac{\theta}{2}|\downarrow\rangle)_1\otimes
 |\downarrow\rangle_2$ are also (orthogonal degenerate)
 eigenstates of the composite
 system at $g=0$, which also lead to zero contribution of the
 second subsystem to the Berry phase in the composite system.
On the other hand, the adiabatic condition in the limit
$g\rightarrow 0$ is $\sim| \omega/(g\sin\theta)|\ll 1$ for
time-independent $\theta$ and $g$, where $\omega=\partial
\phi/\partial t$. This tells that {\it the adiabatic theorem would
break down in the weak coupling limit}. A very small chosen $\omega$
leads the system to difficulties  to finish a cyclic evolution due
to decoherence effects.

Collecting the idea in  \cite{erik04c,fu04,milman06},  we now
explain that the definition of subsystem geometric phase given in
our paper is well defined. To make it clear, we first map the mixed
state of the subsystem into the Bloch ball, then prolong the density
matrix vector until it reaches the unitary Bloch sphere. The two
vectors that represent the two eigenstates of the density matrix
point in opposite directions, respectively. The weighted sum of the
individual PHASES is then understood  as  the weighted sum of the
AREAS  corresponding to the two pure states lying on the unitary
Bloch sphere. As the individual phase is  specified as the area and
the two vectors are a jointed pair, the defined subsystem geometric
phase is gauge invariant and well defined.

In conclusion, we show that in the weak coupling limit $g\rightarrow
0$, the second subsystem makes no contribution to the geometric
phase in  realistic composite systems, instead of $(-\pi)$ in the
comment. As we have emphasized in \cite{yi041}, the definition
$\gamma=\sum_i p_i\gamma_i$ may act as weighted one-particle
geometric phase, it is $U(1)$ gauge invariant as shown in
\cite{fu04} and well defined as we have explained.


\begin{references}
\bibitem{note1} They come  from [1], where the authors take them as
the eigenstates of
the composite system in the limit $g\rightarrow 0$. Note that $\phi$
together with $\theta$ characterize the direction of the magnetic
field which drives only the first subsystem.
\end{references}
\end{document}